\documentclass[fleqn,usenatbib]{mnras}

\usepackage{newtxtext,newtxmath}

\usepackage[T1]{fontenc}
\usepackage{ae,aecompl}

\usepackage{graphicx}	
\usepackage{amsmath}	
\usepackage{amssymb}	

\def\simeq{
\mathrel{\raise.3ex\hbox{$\sim$}\mkern-14mu\lower0.4ex\hbox{$-$}}
}

\def\ltsima{$\; \buildrel < \over \sim \;$}
\def\simlt{\lower.5ex\hbox{\ltsima}}
\def\gtsima{$\; \buildrel > \over \sim \;$}
\def\simgt{\lower.5ex\hbox{\gtsima}}

\def\msun{{\rm M_{\odot}}}

\def\be{\begin{equation}}
\def\ee{\end{equation}}

\def\del#1{{}}
\def\ltsima{$\; \buildrel < \over \sim \;$}
\def\simlt{\lower.5ex\hbox{\ltsima}}
\def\gtsima{$\; \buildrel > \over \sim \;$}
\def\simgt{\lower.5ex\hbox{\gtsima}}

\title[Low--spin SMBHs are more massive]{Slow and fat: low--spin SMBHs are more massive}

\author[K. Zubovas, A. R. King]{Kastytis Zubovas$^{1,2,\star}$ and Andrew King$^{3,4,5}$ \\
  $^{1}$Center for Physical Sciences and Technology, Saul\.{e}tekio al. 3, Vilnius LT-10257, Lithuania\\
  $^{2}$Vilnius University Observatory, Saul\.{e}tekio al. 3, Vilnius LT-10257, Lithuania\\
  $^{3}$Department of Physics \& Astronomy, University of Leicester, Leicester, LE1 7RH, UK \\
  $^{4}$ Astronomical Institute Anton Pannekoek, University of Amsterdam, Science Park 904, 1098 XH Amsterdam, Netherlands\\ 
  $^{5}$ Leiden Observatory, Leiden University, Niels Bohrweg 2, NL-2333 CA Leiden, Netherlands \\
  $^{\star}$ {E-mail:~} {\rm kastytis.zubovas@ftmc.lt} }

\date{Accepted XXX. Received YYY; in original form ZZZ}

\pubyear{2019}

\begin{document}
\label{firstpage}
\pagerange{\pageref{firstpage}--\pageref{lastpage}}
\maketitle

\begin{abstract}
Active galactic nuclei (AGN) probably control the growth of their host galaxies via feedback in the form of wide-angle wind-driven outflows. These establish the observed correlations between supermassive black hole (SMBH) masses and host galaxy properties, e.g. the spheroid velocity dispersion $\sigma$. In this paper we consider the growth of the SMBH once it starts driving a large-scale outflow through the galaxy. To clear the gas and ultimately terminate further growth of both the SMBH and the host galaxy, the black hole must continue to grow its mass significantly, by up to a factor of a few, after reaching this point. The mass increment $\Delta M_{\rm BH}$ depends sensitively on both galaxy size and SMBH spin. The galaxy size dependence leads to $\Delta M_{\rm BH} \propto \sigma^5$ and a steepening of the $M-\sigma$ relation beyond the analytically calculated $M \propto \sigma^4$, in agreement with observation. Slowly--spinning black holes are much less efficient in producing feedback, so at any given $\sigma$ the slowest--spinning black holes should be the most massive. Current observational constraints are consistent with this picture, but insufficient to test it properly; however, this should change with upcoming surveys.

\end{abstract}




\section{Introduction} \label{sec:intro}

It is now generally accepted that most galaxies harbour supermassive black holes (SMBHs) in their centres. During periods of rapid accretion, the SMBHs and their surroundings appear as active galactic nuclei (AGN) that can drive powerful outflows and significantly affect the evolution of the host galaxy \citep{Cicone2015A&A,Fiore2017A&A}. In particular, the mass flow rate in these massive outflows can be several times higher than the star formation rate in the host galaxy \citep{Feruglio2010A&A}. 

The AGN wind-driven outflow model can explain the salient properties of these outflows, as well as their scaling with AGN luminosity \citep{Zubovas2012ApJ}. This model also explains the observed correlation between SMBH masses and the velocity dispersion in the host galaxy \citep[the $M-\sigma$ relation, cf.][]{Kormendy2013ARA&A, McConnell2013ApJ}. In this model, the AGN can only drive large-scale outflows once its luminosity reaches a critical threshold. At this luminosity, the pressure force of the wind produced by the AGN becomes large enough to overcome the weight of the surrounding gas distribution, and gas can be pushed out to arbitrary radii \citep[for a more thorough derivation, see][]{King2010MNRASa}. If we assume that the SMBH at that moment is radiating at a fraction $l$ of its Eddington luminosity, we find a relation for the SMBH mass:
\begin{equation} \label{eq:msigma}
    M_{\rm crit} = \frac{f_{\rm g} \left(1-f_{\rm g}\right)\kappa \sigma^4}{\pi G^2} = 3.09 \times 10^8 \frac{f_{\rm g}}{0.16} \frac{1-f_{\rm g}}{0.84} \frac{\sigma_{200}^4}{l} \, \msun,
\end{equation}
where $\kappa = 0.346$~cm$^2$~g$^{-1}$ is the electron scattering opacity, $\sigma \equiv 200 \sigma_{200}$~km~s$^{-1}$ is the velocity dispersion in the host galaxy spheroid, $G$ is the gravitational constant and $f_{\rm g} \equiv \rho_{\rm g}/\rho_{\rm tot}$ is the gas fraction, i.e. the ratio of gas density $\rho_{\rm g}$ and total density $\rho_{\rm tot}$ in the spheroid. In the above expression, $f_{\rm g}$ is scaled to the cosmological value of $0.16$.

The relation (\ref{eq:msigma}) agrees quite well with observations, even though it has very little freedom in terms of parameter values. The terms involving gas fraction obey $f_{\rm g}\left(1-f_{\rm g}\right) \leq 0.25$, so the actual dependence on $f_{\rm g}$ is weak. The Eddington factor can have a stronger influence, but only if the SMBH maintains a similar $l$ over multiple accretion episodes. To see this, consider that the outflow driven by an AGN affects whatever gas reservoir feeds the black hole. If $l$ is small, the outflow is weak and the reservoir can build up, increasing the accretion rate. Eventually, $l$ approaches unity, and $M_{\rm crit}$ decreases, until the SMBH can efficiently remove most of the gas that might be able to feed it.

One significant disagreement is the slope of the relation, $\alpha$. Observed values are typically higher than $\alpha = 4$, although there is a wide range of values proposed in the literature, ranging from as low as $\alpha \simeq 4.38$ in \cite{Kormendy2013ARA&A} to \cite{McConnell2013ApJ} finding $\alpha \simeq 5.64$. It is important to note that the slope decreases once galaxies are subdivided by morphology: early-type and late-type galaxies have slopes $\alpha_{\rm early} = 5.20$ and $\alpha_{\rm late} = 5.06$, with early-type galaxies having an intercept value twice larger than late-type ones. Similarly, active galaxies (which are less likely to be ellipticals) have a much flatter $M-\sigma$ relation, with a slope of $\alpha \simeq 3.32$ \citep{Xiao2011ApJ}. The picture is further complicated by some SMBHs having masses far above the $M-\sigma$ relation, such as NGC 4889 and NGC 3482 \citep{McConnell2011Natur}. The host galaxies of these SMBHs may have had much higher velocity dispersions in the past \citep{King2019MNRAS}.

Clearly, galaxy size and morphology have a significant impact on the final mass its SMBH achieves. \cite{Zubovas2012MNRASb} suggested a possible explanation: the size of the galaxy determines the energy input required for the outflow to clear the gas out of the galaxy spheroid, and hence the time for which the SMBH must be active after achieving the mass given by eq. (\ref{eq:msigma}). In spiral galaxies, assuming Eddington-limited AGN episodes, this timescale is of order a few Myr, leading to negligible additional growth of the SMBH. In elliptical galaxies, subsequent AGN episodes must last for almost $10^8$~yr, leading to SMBH growth by almost an order of magnitude, producing an offset in the $M-\sigma$ relation between the two populations. If sub-Eddington episodes are assumed, the timescales increase in proportion to $l^{-1}$, but the total SMBH mass change remains the same. Combined with the fact that elliptical galaxies have higher velocity dispersions than spiral galaxies, this leads to an overall steeper $M-\sigma$ relation.

In this paper, we revisit this argument and consider the growth of SMBHs with different spins. We show that the extra mass gained after reaching $M_{\rm crit}$ is proportional to $\eta^{-2}$, where $\eta$ is the spin-dependent radiative efficiency. Therefore, SMBHs with low spins grow to higher final masses than those with high spins, by a factor of a few. Although many other factors contribute to the spread of the $M-\sigma$ relation, a correlation should emerge when deviations from the mean relation are considered in large galaxy samples.

The paper is structured as follows. In Section \ref{sec:driving}, we review the process of driving the gas out of a galaxy and determine the energy input required to do so. In Section \ref{sec:growth}, we calculate the SMBH mass growth during the process of clearing the gas out of the galaxy and suggest the distributions of final SMBH masses that might be expected given some distributions of their spins. In Section \ref{sec:before}, we comment on the influence of SMBH spin on its growth before reaching $M_{\rm crit}$. We discuss and conclude in Section \ref{sec:discuss}.

\section{Driving gas out of the galaxy} \label{sec:driving}

In this section, we recast the argument of \citep{Zubovas2012MNRASb} in terms of energy input into the galactic gas. Once the SMBH reaches its critical mass given in eq. \ref{eq:msigma}, it can begin driving a large-scale outflow throughout the host galaxy. However, in order to prevent further significant SMBH growth, it is necessary that the gas is removed very far from the SMBH. In an elliptical galaxy, this may mean gas removal to the virial radius. Assuming a galaxy  where dark matter and gas are distributed in an isothermal distribution, with gas fraction $f_{\rm g}$, a velocity dispersion $\sigma \equiv \sqrt{GM\left(<R\right)/\left(2R\right)} = 200 \sigma_{200}$~km~s$^{-1}$, and a virial radius $R_{\rm v} = 200 R_{200}$~kpc, the total gass mass within a radius $R$ is
\begin{equation}\label{eq:mass_isotherm}
    M_{\rm g} = f_{\rm g} M\left(<R\right) = \frac{2 f_{\rm g} \sigma^2 R}{G} \sim 5.95\times10^{11} \frac{f_{\rm g}}{0.16} \sigma_{200}^2 R_{200}\, \msun.
\end{equation}
The energy required to unbind this gas is
\begin{equation}\label{eq:energy_isotherm}
    E_{\rm bind} \sim M_{\rm g} \sigma^2 = \frac{2 f_{\rm g} \sigma^4 R}{G} \sim 4.7\times10^{59} \frac{f_{\rm g}}{0.16} \sigma_{200}^4 R_{200}\,{\rm erg}.
\end{equation}
The number does not change significantly if we consider a different density profile. For example, using an NFW \citep{Navarro1997ApJ} profile with concentration parameter $c \equiv R_{\rm v}/a = 10$, where $a$ is the scale radius, gives a binding energy $\sim 40\%$ higher than the isothermal case.

The actual energy that must be injected into the gas in order to shut off further accretion on to the SMBH can differ significantly from the above estimate. It is increased if a lot of gas is dense and can cool down efficiently, and decreased if the gas has significant angular momentum that prevents re-accretion to the centre. Nevertheless, we think the estimate is approximately correct to within an order of magnitude, which is enough for our purposes.

In spiral galaxies, it might be enough to drive the gas out of the bulge, to a distance $R_{\rm b} \equiv 2R_2$~kpc. Subsequently, gas might fall on to the disc before it falls back into the bulge, mix with the disc gas and no longer feed the SMBH. In that case, the gas mass that has to be removed is a factor $R_{\rm b}/R_{\rm v}$ smaller. In an isothermal potential, the ratio of potential energies at radii $r_1$ and $r_2$ is ${\rm ln}\left(r_1/r_2\right)$, so the required energy injection is a factor $\sim \left(R_{\rm v}/R_{\rm b}\right){\rm ln}\left(R_{\rm v}/R_{\rm b}\right) \sim 460$ smaller, i.e.
\begin{equation}\label{eq:energy_bulge}
    E_{\rm bulge} \sim 1.0\times10^{57} \frac{f_{\rm g}}{0.16} \sigma_{200}^4 R_{2}\,{\rm erg}.
\end{equation}

\section{SMBH growth during galaxy clearing} \label{sec:growth}

The energy required to clear the gas out of the galaxy is injected by the AGN, over several activity episodes. The energy supplied by the AGN wind is \citep[cf.][]{King2010MNRASa}
\begin{equation}\label{eq:ewind}
    E_{\rm w} = \frac{\eta}{2}E_{\rm AGN} = \frac{\eta^2}{2}\Delta M_{\rm BH} c^2,
\end{equation}
where $\eta$ is the radiative efficiency of accretion, $E_{\rm AGN}$ is the energy radiated by the AGN and $\Delta M_{\rm BH}$ is the mass growth of the SMBH during the process of galaxy clearing. This energy is absorbed by the gas with a certain efficiency $f < 1$, which depends on the geometry of the gas distribution, the efficiency of gas cooling and the advection of energy beyond the virial radius by the outflowing material. Keeping this efficiency as a free parameter for now, we can equate $E_{\rm bind}$ with $fE_{\rm w}$ to find
\begin{equation}\label{eq:deltam}
    \Delta M_{\rm BH} \sim \frac{2 f_{\rm g} \sigma^4 R}{G} \frac{2}{f \eta^2 c^2} \sim 5.3 \times 10^7 \frac{f_{\rm g}}{0.16} \sigma_{200}^4 R_{200} f^{-1} \eta_{0.1}^{-2} \, \msun.
\end{equation}

The ratio of this mass growth to the $M_{\rm crit}$ value is
\begin{equation}\label{eq:deltamratio}
    \frac{\Delta M_{\rm BH}}{M_{\rm crit}} \sim 0.15 \frac{f_{\rm g}}{0.16} R_{200} f^{-1} \eta_{0.1}^{-2}.
\end{equation}
When calculating this ratio, we assumed that $f_{\rm g} = 0.16$ during the establishment of the critical BH mass, but left $f_{\rm g}$ as a free parameter on larger scales / at later times. 

The equations (\ref{eq:deltam}) and (\ref{eq:deltamratio}) depend on four parameters that may vary significantly among different galaxies. For example, the gas fraction $f_{\rm g}$ may be significantly lower than the cosmological value for a gas-poor galaxy, but may remain at the approximately cosmological value in a gas-rich cluster \citep{Zubovas2012MNRASb}. This leads to cluster galaxies having higher SMBH masses than field galaxies, as observed \citep{McConnell2013ApJ}.

Galaxy sizes, such as $R_{\rm v}$, correlate with velocity dispersion, with $R_{\rm v} \propto \sigma$ \citep[this is one of the projections of the galaxy Fundamental Plane, cf.][]{Djorgovski1987ApJ, Marconi2003ApJ, Cappellari2013MNRAS}. Substituting this relation into eq. (\ref{eq:deltam}) gives $\Delta M_{\rm BH} \propto \sigma^5$, i.e. the $M-\sigma$ relation steepens once the SMBH is able to drive large-scale outflows. There is some indication that such steepening occurs at a particular value of $\sigma$ or corresponding stellar mass \citep{Martin2018arXiv, Krajinovic2018MNRAS}, although it is unclear whether difference in feedback requirements is the driving factor for them.

The coupling efficiency of the wind to the gas, $f$, is presumably rather low. In \cite{Power2011MNRAS}, we used an energy argument similar to the one above to determine that the SMBH would grow by $\sim 40\%$ above $M_{\rm crit}$ while the bulge is being cleared; there we used $R_{\rm v} \sim 400$~kpc, so the result is consistent with eq. (\ref{eq:deltamratio}) assuming $f \simeq 0.75$. This is probably an upper limit, since our argument did not account for the dynamics of the gas, uneven density and other complicating factors. In \cite{Zubovas2012MNRASb}, we showed that in a gas-rich elliptical galaxy, the SMBH may need to grow for $\sim 10^8$~yr in order to clear all the gas out of a galaxy with $R_{\rm v} = 400$~kpc, even assuming Eddington-limited AGN episodes, since most of the injected energy ends up moving gas far beyond the virial radius. Such a long growth period leads to the SMBH growing by $\Delta M_{\rm BH} \simeq 6.5 M_{\rm crit}$; plugging this result into eq. \ref{eq:deltamratio} gives a much lower estimate $f \sim 0.05$. This may be a lower limit, since gas cooling may lead to a narrower and denser outflow \citep[cf.][]{Zubovas2014MNRASa, Richings2018MNRAS, Richings2018MNRASb} and hence better absorption of AGN feedback energy by gas within the virial radius; furthermore, clearing gas out of the virial radius may not be necessary to stop SMBH growth. Evidently, the true value of $f$ is somewhere between these two extremes. We estimate it by the following argument: elliptical galaxies should have black hole masses $M_{\rm BH, el} = M_{\rm crit} + \Delta M_{\rm BH}$, while in spiral galaxies, $M_{\rm BH, sp} \simeq M_{\rm crit}$, due to the much lower energy required to remove gas from the bulge; considering the difference in the intercepts of $M-\sigma$ relation for elliptical and spiral galaxies \citep{McConnell2013ApJ} leads to $\Delta M_{\rm BH} \simeq M_{\rm crit}$ and $f \sim 0.15$.

Finally, and most importantly for the present paper, the radiative efficiency of accretion has a very strong influence on the final SMBH mass. The radiative efficiency of accretion on to a non-spinning (Schwarzschild) black hole is $\eta_{a = 0} = 0.055$, while accretion on to a maximally spinning Kerr black hole releases $\eta_{a = -1} = 0.038$ of rest mass energy if the accretion disc angular momentum is opposite to that of the black hole spin (retrograde case) and $\eta_{a = 1} = 0.42$ if the angular momenta align (prograde case). SMBH growth is composed of many individual episodes lasting $t_{\rm ep} \sim 10^{4}-10^5$~yr \citep{King2015MNRAS, Schawinski2015MNRAS}, each with only a small mass $\Delta M_1 \sim 10^{-3} M_{\rm BH}$ \citep{King2006MNRAS}, which should not affect the value of the SMBH spin significantly. Such events can produce discs stably co- or counter-aligned with the SMBH spin \citep{King2005MNRAS}, so the energy released over many AGN episodes is
\begin{equation}
    E_{\rm AGN} \simeq \left(N_{\rm pr} \eta_{\rm pr} \delta M_{\rm BH, pr} + N_{\rm re} \eta_{\rm re} \delta M_{\rm BH, re}\right) c^2,
\end{equation}
where $N_{\rm pr}$ and $N_{\rm re}$ are the number of episodes where the accretion disc is aligned prograde or retrograde to the SMBH spin, $\eta_{\rm pr}$ and $\eta_{\rm re}$ are the corresponding radiative efficiencies, while $\delta M_{\rm BH, pr}$ and $\delta M_{\rm BH, re}$ are the mass changes in a single prograde or retrograde accretion episode. When $\Delta M_1 \ll M_{\rm BH}$, the probability of prograde and retrograde alignment is approximately the same, and the mass change is similar in both prograde and retrograde cases, giving
\begin{equation}
    E_{\rm AGN} \simeq \frac{\eta_{\rm pr} + \eta_{\rm re}}{2} \delta M_{\rm BH} c^2,
\end{equation}
i.e. the mean radiative efficiency is just the average of the prograde and retrograde cases. The appropriate range of average accretion efficiencies is $0.055 < \langle\eta\rangle < 0.23$, with the maximum value being the mean of the prograde and retrograde accretion efficiencies on to a maximally spinning black hole. Since $\Delta M_{\rm BH} \propto \eta^{-2}$, the value of $\Delta M_{\rm BH}$ can vary by a factor $\sim20$ depending on the SMBH spin. In particular, for a non-spinning SMBH,
\begin{equation}
    \frac{\Delta M_{\rm BH}}{M_{\rm crit}} \sim 0.5 \frac{f_{\rm g}}{0.16} R_{200} f^{-1} \sim 3 \frac{f_{\rm g}}{0.16} R_{200},
\end{equation}
while for a maximally spinning one,
\begin{equation}
    \frac{\Delta M_{\rm BH}}{M_{\rm crit}} \sim 0.028 \frac{f_{\rm g}}{0.16} R_{200} f^{-1} \sim 0.17 \frac{f_{\rm g}}{0.16} R_{200},
\end{equation}
where we used the estimate $f = 0.15$ in the last equality for both cases. If individual accretion events are more likely to align in a prograde fashion \citep{Dotti2013ApJ, Wang2016ApJ}, the average accretion efficiency for accretion on to rapidly spinning SMBHs, and hence the range of possible $\Delta M_{\rm BH}$, becomes even higher.

\begin{figure}
	\includegraphics[width=\columnwidth]{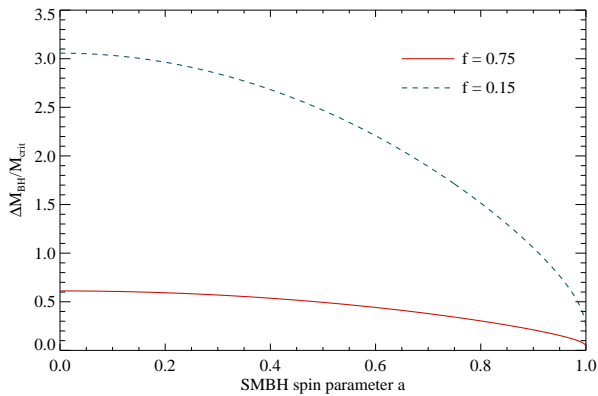}
    \caption{Fractional growth of the SMBH after beginning to drive a large-scale outflow, as function of the SMBH spin parameter $a$. Two lines correspond to different feedback coupling efficiency, $f = 0.75$ is very efficient feedback, $f = 0.15$ is our estimate of a typical value.}
    \label{fig:deltam}
\end{figure}

\begin{figure*}
	\includegraphics[width=0.8\textwidth]{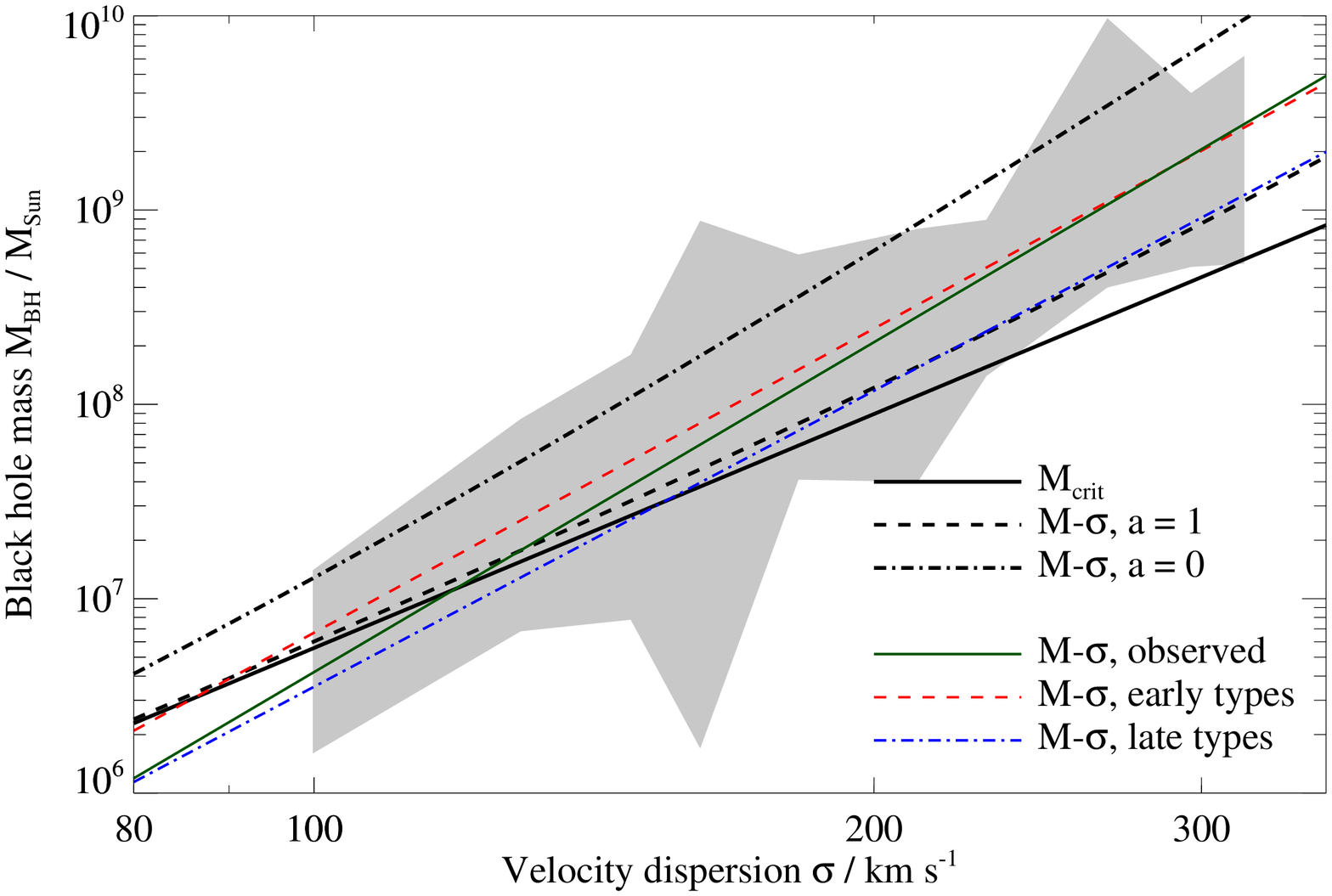}
    \caption{The $M-\sigma$ relation predicted for SMBHs with different spins. Solid black line shows $M_{\rm crit}$ with $f_{\rm g} = 0.05$ (eq. \ref{eq:msigma}), dashed and dot-dashed black lines show the predicted relation for SMBHs with maximal and zero spin, respectively. Green solid line is the observed $M-\sigma$ for the whole sample from \citet{McConnell2013ApJ}, while the red dashed and blue dash-dotted lines show the observed relation for early- and late- type galaxies, respectively. The grey shaded area is the approximate locus of data points from \citet{McConnell2013ApJ}.}
    \label{fig:msigma}
\end{figure*}

We plot the relation $\Delta M_{\rm BH}/M_{\rm crit}$ as a function of $a$ in Figure \ref{fig:deltam} for two possible values of the feedback coupling efficiency: $f = 0.15$ (moderate feedback) and $f = 0.75$ (highly efficient feedback). In both cases, we assume $f_{\rm g} = 0.16$ and $R_{\rm v} = 200$~kpc. For very efficient feedback, the extra SMBH growth is small in all cases, and very precise measurements of SMBH masses are required in order to determine any spin dependence. For less efficient feedback, even knowing the masses to within a factor of 2 is enough to distinguish between fast ($a > 0.9$), medium ($0.65 < a < 0.9$) and slow ($a < 0.65$) rotators.

In Figure \ref{fig:msigma}, we plot the $M-\sigma$ relations that would result from SMBHs having different values of spin. The three black thicker lines show our model predictions: solid line shows $M_{\rm crit}$, dashed line shows $M_{\rm crit} + \Delta M_{\rm BH}$ for $a = 1$ and dot-dashed line shows $M_{\rm crit} + \Delta M_{\rm BH}$ for $a = 0$. When calculating these, we took $f_{\rm g} = 0.05$ to account for the baryon fraction being lower than cosmological in both observed \citep{Dai2010ApJ} and simulated \citep{SantosSantos2016MNRAS} galaxies. We use a relation $R_{\rm v} = 293 \sigma_{200}^{2.19}$~kpc, derived from a combination of the $R_{\rm e} - \sigma$ \citep[where $R_{\rm e}$ is the galaxy effective radius; ][]{Jorgensen2013AJ} and $R_{\rm e} - R_{\rm v}$ \citep{Kravtsov2013ApJ} relations. We also adopt $f = 0.15$. It is clear that galaxies with higher velocity dispersions experience stronger offsets due to their larger virial radii. For comparison, we plot the relations derived by \citet{McConnell2013ApJ} from observations of all (green solid), early-type (red dashed) and late-type (blue dot-dashed) galaxies, and the locus of all data points within their sample (grey shaded region). The relation $M_{\rm crit}$ lies somewhat below the observed relations, especially for the largest galaxies, but the predicted relations for black holes with different spins agree with observations rather well. In particular, the relation for the slowest-spinning SMBHs approximatey traces the upper edge of the locus of data points.

We therefore predict that at any given host galaxy $\sigma$, the most massive black holes have the lowest spins, while the rapidly spinning ones are close to the `average' values given by the $M-\sigma$ relation. More broadly, the residuals of the $M-\sigma$ relation, i.e. the differences between actual SMBH masses and the masses predicted by the $M-\sigma$ relation, should correlate strongly with SMBH spin. Furthermore, given that $R_{200}$ increases with increasing velocity dispersion, we predict that this discrepancy becomes stronger at higher values of $\sigma$ and that if SMBHs were separated into sub-populations by spin, the slow-spinning SMBHs would have a steeper $M-\sigma$ relation slope than rapidly-spinning ones.

\section{SMBH growth before reaching $M_\text{\lowercase{crit}}$} \label{sec:before}

Before the SMBH reaches $M_{\rm crit}$, its growth rate also depends on spin, although less strongly than above $M_{\rm crit}$. The maximum growth rate is
\begin{equation}
    \dot{M}_{\rm BH,max} = \frac{1-\eta}{\eta}\frac{L_{\rm Edd}}{c^2},
\end{equation}
where the factor $1-\eta$ accounts for the loss of mass-energy in the material that falls into the SMBH after radiating a fraction $\eta$ of its mass away. This rate is lower for higher-spin SMBHs: the quantity $\left(1-\eta\right)/\eta$ ranges between $1.4$ for a maximally spinning SMBH accreting from a prograde disc to $17.2$ for a non-spinning SMBH. So a rapidly-spinning SMBH needs a longer period of nuclear activity to reach $M_{\rm crit}$ than a non-spinning one, by as much as a factor $\sim 12$. 
The AGN duty cycle should not depend on the SMBH spin, since it is governed primarily by the gas supply and only becomes affected by feedback significantly once $M_{\rm crit}$ is reached \citep{King2010MNRASa}. Therefore, in a population of galaxies with similar mass observed at a similar redshift, there should be an anti-correlation between SMBH mass and spin. Clearly, this effect is stronger at higher redshift, when fewer black holes have reached $M_{\rm crit}$. More massive galaxies have higher duty cycles \citep{Aversa2015ApJ, Comparat2019arXiv}, hence their SMBHs reach $M_{\rm crit}$ faster. This process may lead to further steepening of the observed $M-\sigma$ relation. It also suggests that small isolated galaxies may be the best locations to search for a correlation between SMBH mass and spin, since the black holes in those galaxies are less affected by mergers.

\section{Discussion and summary}  \label{sec:discuss}

There are few robust estimates of SMBH spins available in the literature, but the available data shows a general trend of more massive black holes spinning more slowly \citep{Brenneman2013brief, Reynolds2013CQGra, Vasudevan2016MNRAS, Reynolds2019NatAs}. Selection effects mean that a lot of low-spin SMBHs are undetected \citep{Vasudevan2016MNRAS}, although it is impossible to predict what mass range they might fall in. However, the relative paucity of rapidly spinning SMBHs with masses $M_{\rm BH} > 10^8 \msun$ suggests that such black holes are very rare. Such a relationship is a natural prediction of our model, where such rapidly-spinning black holes can only exist in rare galaxies with very large velocity dispersions. On the other hand, models where SMBH growth is dominated by mergers would predict some SMBHs to have very large spins \citep{Volonteri2005ApJ}, which should be detectable.

\cite{Xiao2011ApJ} find that the offset from the $M-\sigma$ relation, $\Delta M_{\rm BH}$ anti-correlates with Eddington ratio in AGN. This is consistent with the results of our model. While there are many environmental factors influencing the SMBH accretion rate, the luminosity, and hence the Eddington ratio, also depends on the radiative efficiency, i.e. on SMBH spin. Provided environmental conditions are the same, a black hole with higher spin will have a higher luminosity, and hence a higher Eddington ratio. In our model, these black holes also have lower $\Delta M_{\rm BH}$, consistently with observations. This effect is, of course, degenerate with the fact that it is more difficult to reach a given Eddington factor for a more massive SMBH. Furthermore, $\Delta M_{\rm BH}$ correlates positively with AGN luminosity, suggesting that brighter AGN are powered by more overmassive black holes. This is consistent with our result that the $M-\sigma$ relation steepens once $M_{\rm crit}$ is reached.

The final distribution of SMBH masses depends, among other factors, on the distribution of their spins, which is difficult to constrain via simulations. Some early results suggested that most SMBHs should be spinning at rates close to maximal \citep{Volonteri2005ApJ, Berti2008ApJ, Cao2008MNRAS}, but that spiral galaxies should have lower SMBH spins on average compared with ellipticals \citep{Volonteri2007ApJ}. But accounting for the expected accretion disc masses and the likelihood of stable counter-alignment of the disc and SMBH \citep{King2005MNRAS}, most black holes should be spinning rather slowly \citep{King2008MNRAS}. The expected spin values depend sensitively on whether gas accretion on to the SMBH is chaotic or correlated, the latter producing much higher average spins \citep{Berti2008ApJ, Griffin2018arXiv}. Presently, there is no widespread agreement on the expected distribution of SMBH spins. Observations of black hole mass distributions in galaxies with narrow ranges in $\sigma$, or of distributions of black hole mass residuals from the $M-\sigma$ relation, may help shed some light on the issue: if the spin distribution is flat or bottom--heavy, most black holes have low radiative efficiencies and their masses should have significant offsets from the average $M-\sigma$ relation; on the other hand, if most spins are large, there should be few outliers significantly above the relation. Of course, we do not expect this relation to be clear enough for use in predicting the spins of individual SMBHs. However, a large enough population sample would help determine the broad properties of the spin distribution.

As mentioned in Section \ref{sec:growth} and previously found in \cite{Zubovas2012MNRASb}, there should be a morphological dependence of the $M-\sigma$ relation, elliptical galaxies having much higher offsets and so a potentially higher spread of mass values at a given $\sigma$. This prediction is consistent with the data presented in \cite{Kormendy2013ARA&A}, where masses of SMBHs in classical bulges are generally closer to the $M-\sigma$ relation than of those in elliptical galaxies (see the right panel of their Figure 16). On the other hand, \cite{McConnell2013ApJ} find that the scatter in the $M-\sigma$ relation is higher for late-type galaxies than for early-types, although much of that scatter is caused by galaxies with low-mass SMBHs, which may still be growing.

Another important environmental dependence is the possible correlation between SMBH spin and galaxy mass. Black holes in small galaxies should have experienced less accretion than those in massive galaxies \citep{Nayakshin2009MNRAS, Habouzit2017MNRAS, Yang2018MNRAS, Zubovas2019MNRASa} and probably fewer mergers as well. Both accretion \citep{King2008MNRAS} and mergers \citep{Berti2008ApJ, Gergely2012arXiv} lead to low-to-moderate SMBH spins, so it is plausible that SMBHs in massive galaxies spin slowly. These black holes would then be offset to higher masses from $M_{\rm crit}$ than their counterparts in small galaxies, leading to further steepening of the observed $M-\sigma$ relation. On the other hand, prolonged prograde accretion can spin black holes up to very high rates \citep{Berti2008ApJ, Dotti2013ApJ}; however, in order for this situation to occur, the black hole must always align with the (initially randomly oriented) accretion disc \citep{Scheuer1996MNRAS}; \citet{King2005MNRAS} showed that this is not generally the case. Nevertheless, if such alignment occurs often enough and black holes spin very rapidly, SMBHs in more massive galaxies would generally have smaller values of $\Delta M_{\rm BH}$ than those in smaller galaxies, and the scatter in the $M-\sigma$ relation would show a negative correlation with galaxy mass. \cite{McConnell2013ApJ} found slightly smaller intrinsic scatter in the $M-\sigma$ relation at high velocity dispersions than at low ones, but, intriguingly, the trend is reversed when luminosity or bulge mass is used to distinguish between large and small galaxies. There are other factors that influence the scatter, especially in small galaxies (e.g. tidal perturbations, stochasticity in SMBH feeding and so on), therefore it is difficult to use currently available data to constrain this dependence. In the future, when better measurements of SMBH spin become available, these relations may provide constraints on the SMBH growth history.

There have been various attempts to indirectly estimate SMBH spins, e.g. from the Eddington factor of accretion discs \citep{Piotrovich2016Ap} or jet properties \citep{Kun2013AN, Moscibrodzka2016A&A, Daly2009ApJ, Daly2014MNRAS, Daly2016MNRAS, Daly2019arXiv}. Evolution of the SMBH population over cosmic time also provides constraints on the radiative efficiency, which appears to be low \citep{Soltan1982MNRAS, Merloni2008MNRAS, Davies2019arXiv}, implying low spin values, but this estimate is degenerate with SMBH mass density and/or obscuration \citep{Davies2019arXiv}. We may add another indirect method, related to feedback effects on the host galaxy. If we consider two galaxies with similar values of $\sigma$ and $M_{\rm BH}$, the SMBH with the higher spin will have produced more feedback energy as it grew to its present mass (see eq. \ref{eq:ewind}). Stronger feedback is able to drive larger outflows \citep{Zubovas2012ApJ}, quench star formation in a larger region of the galaxy, and even flatten the central part of the dark matter halo, similarly to supernova explosions in dwarf galaxies \citep{Governato2012MNRAS, Read2019MNRAS}. The integrated feedback effect may not provide a quantitative measure of the SMBH spin, but a qualitative comparison of several galaxies may be possible.

More direct estimates of SMBH spins should come from several new observational instruments and campaigns. LSST and eROSITA may detect stellar transits in front of AGN, which should produce characteristic light curves depending, among other factors, on SMBH spin \citep{Beky2013ApJ}. Detailed observations of AGN accretion disc SEDs with Athena and ground-based optical and UV telescopes would allow determination of the SMBH accretion rate and AGN luminosity, and the ratio of the two gives radiative efficiency which, in turn, determines spin \citep{Dovciak2013arXiv, Padovani2017arXiv}. Athena IFU observations alone might be able to provide SMBH spin constraints with errors $<0.05$ \citep{Barret2019arXiv}. These observatories will also improve the measurements of SMBH masses \citep{Nandra2013arXiv, Dovciak2013arXiv}. Gravitational wave signals can be used to infer the spins of merging black holes \citep{Purrer2016PhRvD}; this might become possible with the launch of the LISA gravitational wave observatory \citep{Sathyaprakash2009LRR, Filloux2012JPhCS}. Over the next decade, our knowledge of SMBH spins should expand considerably \citep{Zoghbi2019BAAS}, to the point where the predictions made in this paper can be definitively tested.

We have investigated the growth of SMBHs once they reach $M_{\rm crit} \simeq 3.1 \times 10^8 \sigma_{200}^4 \, \msun$ and begin driving large-scale outflows in their host galaxies. We showed that the extra mass gained during this epoch, $\Delta M_{\rm BH}$, strongly depends on SMBH spin: slowly--spinning SMBHs gain potentially $20$ times more mass than fast-spinning ones. This effect should lead to an observable differentiation of SMBH masses by spin in galaxies with a given value of $\sigma$. Further, we have $\Delta M_{\rm BH} \propto \sigma^5$ because of the relation between galaxy velocity dispersion and size, steepening the observed $M-\sigma$ relation. These results should be testable in the near future, with upcoming surveys such as 4MOST \citep{deJong2019Msngr} and Athena \citep{Nandra2013arXiv}.

\section*{Acknowledgements}

This work was funded by the Research Council Lithuania grant no. MIP-17-78. Theoretical astrophysics in Leicester is supported by an STFC Consolidated Grant.




\bibliographystyle{mnras}
\bibliography{zubovas} 

\begin{thebibliography}{}
\makeatletter
\relax
\def\mn@urlcharsother{\let\do\@makeother \do\$\do\&\do\#\do\^\do\_\do\%\do\~}
\def\mn@doi{\begingroup\mn@urlcharsother \@ifnextchar [ {\mn@doi@}
  {\mn@doi@[]}}
\def\mn@doi@[#1]#2{\def\@tempa{#1}\ifx\@tempa\@empty \href
  {http://dx.doi.org/#2} {doi:#2}\else \href {http://dx.doi.org/#2} {#1}\fi
  \endgroup}
\def\mn@eprint#1#2{\mn@eprint@#1:#2::\@nil}
\def\mn@eprint@arXiv#1{\href {http://arxiv.org/abs/#1} {{\tt arXiv:#1}}}
\def\mn@eprint@dblp#1{\href {http://dblp.uni-trier.de/rec/bibtex/#1.xml}
  {dblp:#1}}
\def\mn@eprint@#1:#2:#3:#4\@nil{\def\@tempa {#1}\def\@tempb {#2}\def\@tempc
  {#3}\ifx \@tempc \@empty \let \@tempc \@tempb \let \@tempb \@tempa \fi \ifx
  \@tempb \@empty \def\@tempb {arXiv}\fi \@ifundefined
  {mn@eprint@\@tempb}{\@tempb:\@tempc}{\expandafter \expandafter \csname
  mn@eprint@\@tempb\endcsname \expandafter{\@tempc}}}

\bibitem[\protect\citeauthoryear{{Aversa}, {Lapi}, {de Zotti}, {Shankar}  \&
  {Danese}}{{Aversa} et~al.}{2015}]{Aversa2015ApJ}
{Aversa} R.,  {Lapi} A.,  {de Zotti} G.,  {Shankar} F.,   {Danese} L.,  2015,
  \mn@doi [\apj] {10.1088/0004-637X/810/1/74}, \href
  {https://ui.adsabs.harvard.edu/abs/2015ApJ...810...74A} {810, 74}

\bibitem[\protect\citeauthoryear{{Barret} \& {Cappi}}{{Barret} \&
  {Cappi}}{2019}]{Barret2019arXiv}
{Barret} D.,  {Cappi} M.,  2019, arXiv e-prints, \href
  {https://ui.adsabs.harvard.edu/abs/2019arXiv190602734B} {p. arXiv:1906.02734}

\bibitem[\protect\citeauthoryear{{B{\'e}ky} \& {Kocsis}}{{B{\'e}ky} \&
  {Kocsis}}{2013}]{Beky2013ApJ}
{B{\'e}ky} B.,  {Kocsis} B.,  2013, \mn@doi [\apj]
  {10.1088/0004-637X/762/1/35}, \href
  {https://ui.adsabs.harvard.edu/abs/2013ApJ...762...35B} {762, 35}

\bibitem[\protect\citeauthoryear{{Berti} \& {Volonteri}}{{Berti} \&
  {Volonteri}}{2008}]{Berti2008ApJ}
{Berti} E.,  {Volonteri} M.,  2008, \mn@doi [\apj] {10.1086/590379}, \href
  {https://ui.adsabs.harvard.edu/abs/2008ApJ...684..822B} {684, 822}

\bibitem[\protect\citeauthoryear{{Brenneman}}{{Brenneman}}{2013}]{Brenneman2013brief}
{Brenneman} L.,  2013, {Measuring the Angular Momentum of Supermassive Black
  Holes}, \mn@doi{10.1007/978-1-4614-7771-6.
}

\bibitem[\protect\citeauthoryear{{Cao} \& {Li}}{{Cao} \&
  {Li}}{2008}]{Cao2008MNRAS}
{Cao} X.,  {Li} F.,  2008, \mn@doi [\mnras] {10.1111/j.1365-2966.2008.13800.x},
  \href {http://adsabs.harvard.edu/abs/2008MNRAS.390..561C} {390, 561}

\bibitem[\protect\citeauthoryear{{Cappellari} et~al.,}{{Cappellari}
  et~al.}{2013}]{Cappellari2013MNRAS}
{Cappellari} M.,  et~al., 2013, \mn@doi [\mnras] {10.1093/mnras/stt562}, \href
  {https://ui.adsabs.harvard.edu/abs/2013MNRAS.432.1709C} {432, 1709}

\bibitem[\protect\citeauthoryear{{Cicone} et~al.,}{{Cicone}
  et~al.}{2015}]{Cicone2015A&A}
{Cicone} C.,  et~al., 2015, \mn@doi [\aap] {10.1051/0004-6361/201424980}, \href
  {http://adsabs.harvard.edu/abs/2015A%26A...574A..14C} {574, A14}

\bibitem[\protect\citeauthoryear{{Comparat} et~al.,}{{Comparat}
  et~al.}{2019}]{Comparat2019arXiv}
{Comparat} J.,  et~al., 2019, arXiv e-prints, \href
  {https://ui.adsabs.harvard.edu/abs/2019arXiv190110866C} {p. arXiv:1901.10866}

\bibitem[\protect\citeauthoryear{{Dai}, {Bregman}, {Kochanek}  \&
  {Rasia}}{{Dai} et~al.}{2010}]{Dai2010ApJ}
{Dai} X.,  {Bregman} J.~N.,  {Kochanek} C.~S.,   {Rasia} E.,  2010, \mn@doi
  [\apj] {10.1088/0004-637X/719/1/119}, \href
  {https://ui.adsabs.harvard.edu/abs/2010ApJ...719..119D} {719, 119}

\bibitem[\protect\citeauthoryear{{Daly}}{{Daly}}{2009}]{Daly2009ApJ}
{Daly} R.~A.,  2009, \mn@doi [\apjl] {10.1088/0004-637X/696/1/L32}, \href
  {https://ui.adsabs.harvard.edu/abs/2009ApJ...696L..32D} {696, L32}

\bibitem[\protect\citeauthoryear{{Daly}}{{Daly}}{2016}]{Daly2016MNRAS}
{Daly} R.~A.,  2016, \mn@doi [\mnras] {10.1093/mnrasl/slw010}, \href
  {https://ui.adsabs.harvard.edu/abs/2016MNRAS.458L..24D} {458, L24}

\bibitem[\protect\citeauthoryear{{Daly}}{{Daly}}{2019}]{Daly2019arXiv}
{Daly} R.~A.,  2019, arXiv e-prints, \href
  {https://ui.adsabs.harvard.edu/abs/2019arXiv190511319D} {p. arXiv:1905.11319}

\bibitem[\protect\citeauthoryear{{Daly} \& {Sprinkle}}{{Daly} \&
  {Sprinkle}}{2014}]{Daly2014MNRAS}
{Daly} R.~A.,  {Sprinkle} T.~B.,  2014, \mn@doi [\mnras]
  {10.1093/mnras/stt2433}, \href
  {https://ui.adsabs.harvard.edu/abs/2014MNRAS.438.3233D} {438, 3233}

\bibitem[\protect\citeauthoryear{{Davies}, {Hennawi}  \& {Eilers}}{{Davies}
  et~al.}{2019}]{Davies2019arXiv}
{Davies} F.~B.,  {Hennawi} J.~F.,   {Eilers} A.-C.,  2019, arXiv e-prints,
  \href {https://ui.adsabs.harvard.edu/abs/2019arXiv190610130D} {p.
  arXiv:1906.10130}

\bibitem[\protect\citeauthoryear{{Djorgovski} \& {Davis}}{{Djorgovski} \&
  {Davis}}{1987}]{Djorgovski1987ApJ}
{Djorgovski} S.,  {Davis} M.,  1987, \mn@doi [\apj] {10.1086/164948}, \href
  {http://adsabs.harvard.edu/abs/1987ApJ...313...59D} {313, 59}

\bibitem[\protect\citeauthoryear{{Dotti}, {Colpi}, {Pallini}, {Perego}  \&
  {Volonteri}}{{Dotti} et~al.}{2013}]{Dotti2013ApJ}
{Dotti} M.,  {Colpi} M.,  {Pallini} S.,  {Perego} A.,   {Volonteri} M.,  2013,
  \mn@doi [\apj] {10.1088/0004-637X/762/2/68}, \href
  {https://ui.adsabs.harvard.edu/abs/2013ApJ...762...68D} {762, 68}

\bibitem[\protect\citeauthoryear{{Dovciak} et~al.,}{{Dovciak}
  et~al.}{2013}]{Dovciak2013arXiv}
{Dovciak} M.,  et~al., 2013, arXiv e-prints, \href
  {https://ui.adsabs.harvard.edu/abs/2013arXiv1306.2331D} {p. arXiv:1306.2331}

\bibitem[\protect\citeauthoryear{{Feruglio}, {Maiolino}, {Piconcelli}, {Menci},
  {Aussel}, {Lamastra}  \& {Fiore}}{{Feruglio} et~al.}{2010}]{Feruglio2010A&A}
{Feruglio} C.,  {Maiolino} R.,  {Piconcelli} E.,  {Menci} N.,  {Aussel} H.,
  {Lamastra} A.,   {Fiore} F.,  2010, \mn@doi [\aap]
  {10.1051/0004-6361/201015164}, \href
  {http://adsabs.harvard.edu/abs/2010A%26A...518L.155F} {518, L155+}

\bibitem[\protect\citeauthoryear{{Filloux}, {de Preitas Pacheco}, {Durier}  \&
  {de Araujo}}{{Filloux} et~al.}{2012}]{Filloux2012JPhCS}
{Filloux} C.,  {de Preitas Pacheco} J.~A.,  {Durier} F.,   {de Araujo}
  J.~C.~N.,  2012, in Journal of Physics Conference Series. p. 012046,
  \mn@doi{10.1088/1742-6596/363/1/012046}

\bibitem[\protect\citeauthoryear{{Fiore} et~al.,}{{Fiore}
  et~al.}{2017}]{Fiore2017A&A}
{Fiore} F.,  et~al., 2017, \mn@doi [\aap] {10.1051/0004-6361/201629478}, \href
  {http://adsabs.harvard.edu/abs/2017A%26A...601A.143F} {601, A143}

\bibitem[\protect\citeauthoryear{{Gergely} \& {Biermann}}{{Gergely} \&
  {Biermann}}{2012}]{Gergely2012arXiv}
{Gergely} L.~{\'A}.,  {Biermann} P.~L.,  2012, arXiv e-prints, \href
  {https://ui.adsabs.harvard.edu/abs/2012arXiv1208.5251G} {p. arXiv:1208.5251}

\bibitem[\protect\citeauthoryear{{Governato} et~al.,}{{Governato}
  et~al.}{2012}]{Governato2012MNRAS}
{Governato} F.,  et~al., 2012, \mn@doi [\mnras]
  {10.1111/j.1365-2966.2012.20696.x}, \href
  {http://adsabs.harvard.edu/abs/2012MNRAS.422.1231G} {422, 1231}

\bibitem[\protect\citeauthoryear{{Griffin}, {Lacey}, {Gonzalez-Perez}, {Lagos},
  {Baugh}  \& {Fanidakis}}{{Griffin} et~al.}{2018}]{Griffin2018arXiv}
{Griffin} A.~J.,  {Lacey} C.~G.,  {Gonzalez-Perez} V.,  {Lagos} C. d.~P.,
  {Baugh} C.~M.,   {Fanidakis} N.,  2018, arXiv e-prints, \href
  {https://ui.adsabs.harvard.edu/abs/2018arXiv180608370G} {p. arXiv:1806.08370}

\bibitem[\protect\citeauthoryear{{Habouzit}, {Volonteri}  \&
  {Dubois}}{{Habouzit} et~al.}{2017}]{Habouzit2017MNRAS}
{Habouzit} M.,  {Volonteri} M.,   {Dubois} Y.,  2017, \mn@doi [\mnras]
  {10.1093/mnras/stx666}, \href
  {http://adsabs.harvard.edu/abs/2017MNRAS.468.3935H} {468, 3935}

\bibitem[\protect\citeauthoryear{{J{\o}rgensen} \& {Chiboucas}}{{J{\o}rgensen}
  \& {Chiboucas}}{2013}]{Jorgensen2013AJ}
{J{\o}rgensen} I.,  {Chiboucas} K.,  2013, \mn@doi [\aj]
  {10.1088/0004-6256/145/3/77}, \href
  {https://ui.adsabs.harvard.edu/abs/2013AJ....145...77J} {145, 77}

\bibitem[\protect\citeauthoryear{{King}}{{King}}{2010}]{King2010MNRASa}
{King} A.~R.,  2010, \mn@doi [\mnras] {10.1111/j.1365-2966.2009.16013.x}, \href
  {http://adsabs.harvard.edu/abs/2010MNRAS.402.1516K} {402, 1516}

\bibitem[\protect\citeauthoryear{{King} \& {Nealon}}{{King} \&
  {Nealon}}{2019}]{King2019MNRAS}
{King} A.,  {Nealon} R.,  2019, \mn@doi [\mnras] {10.1093/mnras/stz1569}, \href
  {https://ui.adsabs.harvard.edu/abs/2019MNRAS.487.4827K} {487, 4827}

\bibitem[\protect\citeauthoryear{{King} \& {Nixon}}{{King} \&
  {Nixon}}{2015}]{King2015MNRAS}
{King} A.,  {Nixon} C.,  2015, \mn@doi [\mnras] {10.1093/mnrasl/slv098}, \href
  {http://adsabs.harvard.edu/abs/2015MNRAS.453L..46K} {453, L46}

\bibitem[\protect\citeauthoryear{{King} \& {Pringle}}{{King} \&
  {Pringle}}{2006}]{King2006MNRAS}
{King} A.~R.,  {Pringle} J.~E.,  2006, \mn@doi [\mnras]
  {10.1111/j.1745-3933.2006.00249.x}, \href
  {http://adsabs.harvard.edu/abs/2006MNRAS.373L..90K} {373, L90}

\bibitem[\protect\citeauthoryear{{King}, {Lubow}, {Ogilvie}  \&
  {Pringle}}{{King} et~al.}{2005}]{King2005MNRAS}
{King} A.~R.,  {Lubow} S.~H.,  {Ogilvie} G.~I.,   {Pringle} J.~E.,  2005,
  \mn@doi [\mnras] {10.1111/j.1365-2966.2005.09378.x}, \href
  {http://adsabs.harvard.edu/abs/2005MNRAS.363...49K} {363, 49}

\bibitem[\protect\citeauthoryear{{King}, {Pringle}  \& {Hofmann}}{{King}
  et~al.}{2008}]{King2008MNRAS}
{King} A.~R.,  {Pringle} J.~E.,   {Hofmann} J.~A.,  2008, \mn@doi [\mnras]
  {10.1111/j.1365-2966.2008.12943.x}, \href
  {http://adsabs.harvard.edu/abs/2008MNRAS.385.1621K} {385, 1621}

\bibitem[\protect\citeauthoryear{{Kormendy} \& {Ho}}{{Kormendy} \&
  {Ho}}{2013}]{Kormendy2013ARA&A}
{Kormendy} J.,  {Ho} L.~C.,  2013, \mn@doi [\araa]
  {10.1146/annurev-astro-082708-101811}, \href
  {http://adsabs.harvard.edu/abs/2013ARA%26A..51..511K} {51, 511}

\bibitem[\protect\citeauthoryear{{Krajnovi{\'c}}, {Cappellari}  \&
  {McDermid}}{{Krajnovi{\'c}} et~al.}{2018}]{Krajinovic2018MNRAS}
{Krajnovi{\'c}} D.,  {Cappellari} M.,   {McDermid} R.~M.,  2018, \mn@doi
  [\mnras] {10.1093/mnras/stx2704}, \href
  {https://ui.adsabs.harvard.edu/abs/2018MNRAS.473.5237K} {473, 5237}

\bibitem[\protect\citeauthoryear{{Kravtsov}}{{Kravtsov}}{2013}]{Kravtsov2013ApJ}
{Kravtsov} A.~V.,  2013, \mn@doi [\apjl] {10.1088/2041-8205/764/2/L31}, \href
  {https://ui.adsabs.harvard.edu/abs/2013ApJ...764L..31K} {764, L31}

\bibitem[\protect\citeauthoryear{{Kun}, {Wiita}, {Gergely}, {Keresztes},
  {Gopal-Krishna}  \& {Biermann}}{{Kun} et~al.}{2013}]{Kun2013AN}
{Kun} E.,  {Wiita} P.~J.,  {Gergely} L.~{\'A}.,  {Keresztes} Z.,
  {Gopal-Krishna}  {Biermann} P.~L.,  2013, \mn@doi [Astronomische Nachrichten]
  {10.1002/asna.201211986}, \href
  {https://ui.adsabs.harvard.edu/abs/2013AN....334.1024K} {334, 1024}

\bibitem[\protect\citeauthoryear{{Marconi} \& {Hunt}}{{Marconi} \&
  {Hunt}}{2003}]{Marconi2003ApJ}
{Marconi} A.,  {Hunt} L.~K.,  2003, \mn@doi [\apjl] {10.1086/375804}, \href
  {http://adsabs.harvard.edu/abs/2003ApJ...589L..21M} {589, L21}

\bibitem[\protect\citeauthoryear{{Martin-Navarro} \& {Mezcua}}{{Martin-Navarro}
  \& {Mezcua}}{2018}]{Martin2018arXiv}
{Martin-Navarro} I.,  {Mezcua} M.,  2018, preprint, \href
  {http://adsabs.harvard.edu/abs/2018arXiv180207277M} {} (\mn@eprint {arXiv}
  {1802.07277})

\bibitem[\protect\citeauthoryear{{McConnell} \& {Ma}}{{McConnell} \&
  {Ma}}{2013}]{McConnell2013ApJ}
{McConnell} N.~J.,  {Ma} C.-P.,  2013, \mn@doi [\apj]
  {10.1088/0004-637X/764/2/184}, \href
  {http://adsabs.harvard.edu/abs/2013ApJ...764..184M} {764, 184}

\bibitem[\protect\citeauthoryear{{McConnell}, {Ma}, {Gebhardt}, {Wright},
  {Murphy}, {Lauer}, {Graham}  \& {Richstone}}{{McConnell}
  et~al.}{2011}]{McConnell2011Natur}
{McConnell} N.~J.,  {Ma} C.-P.,  {Gebhardt} K.,  {Wright} S.~A.,  {Murphy}
  J.~D.,  {Lauer} T.~R.,  {Graham} J.~R.,   {Richstone} D.~O.,  2011, \nat,
  \href {http://adsabs.harvard.edu/abs/2011Natur.480..215M} {480, 215}

\bibitem[\protect\citeauthoryear{{Merloni} \& {Heinz}}{{Merloni} \&
  {Heinz}}{2008}]{Merloni2008MNRAS}
{Merloni} A.,  {Heinz} S.,  2008, \mn@doi [\mnras]
  {10.1111/j.1365-2966.2008.13472.x}, \href
  {http://adsabs.harvard.edu/abs/2008MNRAS.388.1011M} {388, 1011}

\bibitem[\protect\citeauthoryear{{Mo{\'s}cibrodzka}, {Falcke}  \&
  {Noble}}{{Mo{\'s}cibrodzka} et~al.}{2016}]{Moscibrodzka2016A&A}
{Mo{\'s}cibrodzka} M.,  {Falcke} H.,   {Noble} S.,  2016, \mn@doi [\aap]
  {10.1051/0004-6361/201629157}, \href
  {https://ui.adsabs.harvard.edu/abs/2016A&A...596A..13M} {596, A13}

\bibitem[\protect\citeauthoryear{{Nandra} et~al.,}{{Nandra}
  et~al.}{2013}]{Nandra2013arXiv}
{Nandra} K.,  et~al., 2013, arXiv e-prints, \href
  {https://ui.adsabs.harvard.edu/abs/2013arXiv1306.2307N} {p. arXiv:1306.2307}

\bibitem[\protect\citeauthoryear{{Navarro}, {Frenk}  \& {White}}{{Navarro}
  et~al.}{1997}]{Navarro1997ApJ}
{Navarro} J.~F.,  {Frenk} C.~S.,   {White} S.~D.~M.,  1997, \mn@doi [\apj]
  {10.1086/304888}, \href {http://adsabs.harvard.edu/abs/1997ApJ...490..493N}
  {490, 493}

\bibitem[\protect\citeauthoryear{{Nayakshin}, {Wilkinson}  \&
  {King}}{{Nayakshin} et~al.}{2009}]{Nayakshin2009MNRAS}
{Nayakshin} S.,  {Wilkinson} M.~I.,   {King} A.,  2009, \mn@doi [\mnras]
  {10.1111/j.1745-3933.2009.00709.x}, \href
  {http://adsabs.harvard.edu/abs/2009MNRAS.398L..54N} {398, L54}

\bibitem[\protect\citeauthoryear{{Padovani} et~al.,}{{Padovani}
  et~al.}{2017}]{Padovani2017arXiv}
{Padovani} P.,  et~al., 2017, arXiv e-prints, \href
  {https://ui.adsabs.harvard.edu/abs/2017arXiv170506064P} {p. arXiv:1705.06064}

\bibitem[\protect\citeauthoryear{{Piotrovich}, {Buliga}, {Gnedin}, {Mikhailov}
  \& {Natsvlishvili}}{{Piotrovich} et~al.}{2016}]{Piotrovich2016Ap}
{Piotrovich} M.~Y.,  {Buliga} S.~D.,  {Gnedin} Y.~N.,  {Mikhailov} A.~G.,
  {Natsvlishvili} T.~M.,  2016, \mn@doi [Astrophysics]
  {10.1007/s10511-016-9447-4}, \href
  {https://ui.adsabs.harvard.edu/abs/2016Ap.....59..439P} {59, 439}

\bibitem[\protect\citeauthoryear{{Power}, {Zubovas}, {Nayakshin}  \&
  {King}}{{Power} et~al.}{2011}]{Power2011MNRAS}
{Power} C.,  {Zubovas} K.,  {Nayakshin} S.,   {King} A.~R.,  2011, \mn@doi
  [\mnras] {10.1111/j.1745-3933.2011.01048.x}, \href
  {http://adsabs.harvard.edu/abs/2011MNRAS.413L.110P} {413, L110}

\bibitem[\protect\citeauthoryear{{P{\"u}rrer}, {Hannam}  \&
  {Ohme}}{{P{\"u}rrer} et~al.}{2016}]{Purrer2016PhRvD}
{P{\"u}rrer} M.,  {Hannam} M.,   {Ohme} F.,  2016, \mn@doi [\prd]
  {10.1103/PhysRevD.93.084042}, \href
  {https://ui.adsabs.harvard.edu/abs/2016PhRvD..93h4042P} {93, 084042}

\bibitem[\protect\citeauthoryear{{Read}, {Walker}  \& {Steger}}{{Read}
  et~al.}{2019}]{Read2019MNRAS}
{Read} J.~I.,  {Walker} M.~G.,   {Steger} P.,  2019, \mn@doi [\mnras]
  {10.1093/mnras/sty3404}, \href
  {https://ui.adsabs.harvard.edu/abs/2019MNRAS.484.1401R} {484, 1401}

\bibitem[\protect\citeauthoryear{{Reynolds}}{{Reynolds}}{2013}]{Reynolds2013CQGra}
{Reynolds} C.~S.,  2013, \mn@doi [Classical and Quantum Gravity]
  {10.1088/0264-9381/30/24/244004}, \href
  {https://ui.adsabs.harvard.edu/abs/2013CQGra..30x4004R} {30, 244004}

\bibitem[\protect\citeauthoryear{{Reynolds}}{{Reynolds}}{2019}]{Reynolds2019NatAs}
{Reynolds} C.~S.,  2019, \mn@doi [Nature Astronomy]
  {10.1038/s41550-018-0665-z}, \href
  {https://ui.adsabs.harvard.edu/abs/2019NatAs...3...41R} {3, 41}

\bibitem[\protect\citeauthoryear{{Richings} \&
  {Faucher-Gigu{\`e}re}}{{Richings} \&
  {Faucher-Gigu{\`e}re}}{2018a}]{Richings2018MNRAS}
{Richings} A.~J.,  {Faucher-Gigu{\`e}re} C.-A.,  2018a, \mn@doi [\mnras]
  {10.1093/mnras/stx3014}, \href
  {http://adsabs.harvard.edu/abs/2018MNRAS.474.3673R} {474, 3673}

\bibitem[\protect\citeauthoryear{{Richings} \&
  {Faucher-Gigu{\`e}re}}{{Richings} \&
  {Faucher-Gigu{\`e}re}}{2018b}]{Richings2018MNRASb}
{Richings} A.~J.,  {Faucher-Gigu{\`e}re} C.-A.,  2018b, \mn@doi [\mnras]
  {10.1093/mnras/sty1285}, \href
  {http://adsabs.harvard.edu/abs/2018MNRAS.478.3100R} {478, 3100}

\bibitem[\protect\citeauthoryear{{Santos-Santos}, {Brook}, {Stinson}, {Di
  Cintio}, {Wadsley}, {Dom{\'\i}nguez-Tenreiro}, {Gottl{\"o}ber}  \&
  {Yepes}}{{Santos-Santos} et~al.}{2016}]{SantosSantos2016MNRAS}
{Santos-Santos} I.~M.,  {Brook} C.~B.,  {Stinson} G.,  {Di Cintio} A.,
  {Wadsley} J.,  {Dom{\'\i}nguez-Tenreiro} R.,  {Gottl{\"o}ber} S.,   {Yepes}
  G.,  2016, \mn@doi [\mnras] {10.1093/mnras/stv2335}, \href
  {https://ui.adsabs.harvard.edu/abs/2016MNRAS.455..476S} {455, 476}

\bibitem[\protect\citeauthoryear{{Sathyaprakash} \& {Schutz}}{{Sathyaprakash}
  \& {Schutz}}{2009}]{Sathyaprakash2009LRR}
{Sathyaprakash} B.~S.,  {Schutz} B.~F.,  2009, \mn@doi [Living Reviews in
  Relativity] {10.12942/lrr-2009-2}, \href
  {https://ui.adsabs.harvard.edu/abs/2009LRR....12....2S} {12, 2}

\bibitem[\protect\citeauthoryear{{Schawinski}, {Koss}, {Berney}  \&
  {Sartori}}{{Schawinski} et~al.}{2015}]{Schawinski2015MNRAS}
{Schawinski} K.,  {Koss} M.,  {Berney} S.,   {Sartori} L.~F.,  2015, \mn@doi
  [\mnras] {10.1093/mnras/stv1136}, \href
  {http://adsabs.harvard.edu/abs/2015MNRAS.451.2517S} {451, 2517}

\bibitem[\protect\citeauthoryear{{Scheuer} \& {Feiler}}{{Scheuer} \&
  {Feiler}}{1996}]{Scheuer1996MNRAS}
{Scheuer} P.~A.~G.,  {Feiler} R.,  1996, \mn@doi [\mnras]
  {10.1093/mnras/282.1.291}, \href
  {https://ui.adsabs.harvard.edu/abs/1996MNRAS.282..291S} {282, 291}

\bibitem[\protect\citeauthoryear{{Soltan}}{{Soltan}}{1982}]{Soltan1982MNRAS}
{Soltan} A.,  1982, \mnras, \href
  {http://adsabs.harvard.edu/abs/1982MNRAS.200..115S} {200, 115}

\bibitem[\protect\citeauthoryear{{Vasudevan}, {Fabian}, {Reynolds}, {Aird},
  {Dauser}  \& {Gallo}}{{Vasudevan} et~al.}{2016}]{Vasudevan2016MNRAS}
{Vasudevan} R.~V.,  {Fabian} A.~C.,  {Reynolds} C.~S.,  {Aird} J.,  {Dauser}
  T.,   {Gallo} L.~C.,  2016, \mn@doi [\mnras] {10.1093/mnras/stw363}, \href
  {https://ui.adsabs.harvard.edu/abs/2016MNRAS.458.2012V} {458, 2012}

\bibitem[\protect\citeauthoryear{{Volonteri}, {Madau}, {Quataert}  \&
  {Rees}}{{Volonteri} et~al.}{2005}]{Volonteri2005ApJ}
{Volonteri} M.,  {Madau} P.,  {Quataert} E.,   {Rees} M.~J.,  2005, \mn@doi
  [\apj] {10.1086/426858}, \href
  {https://ui.adsabs.harvard.edu/abs/2005ApJ...620...69V} {620, 69}

\bibitem[\protect\citeauthoryear{{Volonteri}, {Sikora}  \&
  {Lasota}}{{Volonteri} et~al.}{2007}]{Volonteri2007ApJ}
{Volonteri} M.,  {Sikora} M.,   {Lasota} J.-P.,  2007, \mn@doi [\apj]
  {10.1086/521186}, \href
  {https://ui.adsabs.harvard.edu/abs/2007ApJ...667..704V} {667, 704}

\bibitem[\protect\citeauthoryear{{Wang}, {Xu}, {Xu}  \& {Wei}}{{Wang}
  et~al.}{2016}]{Wang2016ApJ}
{Wang} J.,  {Xu} Y.,  {Xu} D.~W.,   {Wei} J.~Y.,  2016, \mn@doi [\apj]
  {10.3847/2041-8205/833/1/L2}, \href
  {https://ui.adsabs.harvard.edu/abs/2016ApJ...833L...2W} {833, L2}

\bibitem[\protect\citeauthoryear{{Xiao}, {Barth}, {Greene}, {Ho}, {Bentz},
  {Ludwig}  \& {Jiang}}{{Xiao} et~al.}{2011}]{Xiao2011ApJ}
{Xiao} T.,  {Barth} A.~J.,  {Greene} J.~E.,  {Ho} L.~C.,  {Bentz} M.~C.,
  {Ludwig} R.~R.,   {Jiang} Y.,  2011, \mn@doi [\apj]
  {10.1088/0004-637X/739/1/28}, \href
  {https://ui.adsabs.harvard.edu/abs/2011ApJ...739...28X} {739, 28}

\bibitem[\protect\citeauthoryear{{Yang} et~al.,}{{Yang}
  et~al.}{2018}]{Yang2018MNRAS}
{Yang} G.,  et~al., 2018, \mn@doi [\mnras] {10.1093/mnras/stx2805}, \href
  {http://adsabs.harvard.edu/abs/2018MNRAS.475.1887Y} {475, 1887}

\bibitem[\protect\citeauthoryear{{Zoghbi} et~al.,}{{Zoghbi}
  et~al.}{2019}]{Zoghbi2019BAAS}
{Zoghbi} A.,  et~al., 2019, in \baas. p.~62 (\mn@eprint {arXiv} {1903.05469})

\bibitem[\protect\citeauthoryear{{Zubovas}}{{Zubovas}}{2019}]{Zubovas2019MNRASa}
{Zubovas} K.,  2019, \mn@doi [\mnras] {10.1093/mnras/sty3211}, \href
  {https://ui.adsabs.harvard.edu/abs/2019MNRAS.483.1957Z} {483, 1957}

\bibitem[\protect\citeauthoryear{{Zubovas} \& {King}}{{Zubovas} \&
  {King}}{2012a}]{Zubovas2012MNRASb}
{Zubovas} K.,  {King} A.~R.,  2012a, \mn@doi [\mnras]
  {10.1111/j.1365-2966.2012.21845.x}, \href
  {http://adsabs.harvard.edu/abs/2012MNRAS.426.2751Z} {426, 2751}

\bibitem[\protect\citeauthoryear{{Zubovas} \& {King}}{{Zubovas} \&
  {King}}{2012b}]{Zubovas2012ApJ}
{Zubovas} K.,  {King} A.,  2012b, \mn@doi [\apjl]
  {10.1088/2041-8205/745/2/L34}, \href
  {http://adsabs.harvard.edu/abs/2012ApJ...745L..34Z} {745, L34}

\bibitem[\protect\citeauthoryear{{Zubovas} \& {King}}{{Zubovas} \&
  {King}}{2014}]{Zubovas2014MNRASa}
{Zubovas} K.,  {King} A.~R.,  2014, \mn@doi [\mnras] {10.1093/mnras/stt2472},
  \href {http://adsabs.harvard.edu/abs/2014MNRAS.439..400Z} {439, 400}

\bibitem[\protect\citeauthoryear{{de Jong} et~al.,}{{de Jong}
  et~al.}{2019}]{deJong2019Msngr}
{de Jong} R.~S.,  et~al., 2019, \mn@doi [The Messenger]
  {10.18727/0722-6691/5117}, \href
  {https://ui.adsabs.harvard.edu/abs/2019Msngr.175....3D} {175, 3}

\makeatother
\end{thebibliography}








\bsp	
\label{lastpage}
\end{document}